\documentclass[12pt]{article}

\usepackage{amsmath}
\usepackage{graphicx,psfrag,epsf}
\usepackage{enumerate}
\usepackage{natbib}
\usepackage{url} 
\usepackage{kotex}
\usepackage{multirow}
\usepackage[small]{caption}
\usepackage{rotating}

\usepackage{color}
\usepackage{amssymb}
\usepackage{rotating}
\usepackage[multiple]{footmisc}

\usepackage[linesnumbered,ruled]{algorithm2e}

\newcommand{\blind}{0}

\usepackage{epsfig}
\usepackage{color}
\usepackage{comment}
\usepackage{ulem}
\usepackage{soul}

\newtheorem{prop}{Proposition}
\newtheorem{assump}{Assumption}
\newtheorem{thm}{Theorem}

\def\calX{{\ensuremath{\mathcal X}}}
\def\calT{{\ensuremath{\mathcal T}}}
\def\calA{{\ensuremath{\mathcal A}}}

\begin{document}


\def\spacingset#1{\renewcommand{\baselinestretch}%
	{#1}\small\normalsize} \spacingset{1}


\if0\blind
{
	\title{\bf Propensity Process: a Balancing Functional}
	\author{Pallavi S. Mishra-Kalyani\\
		Department of Biostatistics and Bioinformatics\\
Emory University		\vspace{0.05in}\\
		Brent A. Johnson\\
		Department of Biostatistics and Computational Biology\\
		University of Rochester\\
		and \\
		Qi Long\footnotemark[1]\\
		Department of Biostatistics, Epidemiology, and Informatics\\
		University of Pennsylvania}
	\date{}
	\maketitle
} \fi

\if1\blind
{
	\bigskip
	\bigskip
	\bigskip
	\begin{center}
		{\LARGE\bf  Propensity Process: a Balancing Functional}
	\end{center}
	\medskip
} \fi

\bigskip
\begin{abstract}
	In observational clinic registries, time to treatment is often of interest, but treatment can be given at any time during follow-up and there is no structure or intervention to ensure regular clinic visits for data collection. To address these challenges, we introduce the time-dependent propensity process as a generalization of the propensity score.  We show that the propensity process balances the entire time-varying covariate history which cannot be achieved by existing propensity score methods and that treatment assignment is strongly ignorable conditional on the propensity process. We develop methods for estimating the propensity process using observed data and for matching based on the propensity process. We illustrate the propensity process method using the Emory Amyotrophic Lateral Sclerosis (ALS) Registry data.	
\end{abstract}

\noindent%
\vfill

\newpage
\spacingset{1.45} 

\section{Introduction} \label{sec:intro}

Amyotrophic lateral sclerosis (ALS) is a rare progressive neurological disorder resulting in the degeneration of both upper motor neurons of the cerebral cortex and lower motor neurons of the spinal cord and peripheral nervous system, with a very poor prognosis.  Currently, there is no cure for ALS and clinical care is generally limited to treating secondary infections and palliative care, such as surgically inserting a percutaneous endogastrostomy (PEG) tube to provide enteral nutrition for individuals having difficulty swallowing \citep{procaccini2008}. Our objective is to assess the effect of inserting a PEG feeding tube on preventing weight loss. PEG insertion is an individual decision and one that must be made while the individual is strong enough to proceed with surgery.  Hence, a randomized controlled trial to study the effect of PEG would be implausible.  We develop new methods to evaluate PEG using data from the Emory ALS Clinic registry.

Let $T$ denote the continuously-defined time of PEG insertion for a randomly selected patient from the population.  The observed outcome $Y$ is collected at or just after a fixed point at time $L$, which consequently restricts the time of PEG insertion. If subjects were randomly assigned to receive PEG prior to $L$ and randomly assigned to treatment times, then both treatment effect and dose-response curve could be estimated using standard methods. However, treatment assignment depends on patient characteristics and confounds the effect of treatment on outcome.  To remove confounding associated with covariate imbalance among treatment levels, we rely on the general concept of the propensity score \citep{rosenbaumandrubin1983,rubin1996}.

When treatment assignment is binary, the propensity score \citep{rosenbaumandrubin1983} is defined as probability of receiving a treatment given a set of observed variables. Generalizations of the propensity score as a balancing score have been investigated in various settings \citep{hirano2004,imaiandvandyk2004,hansen2008prognostic,allen2011control,hu2014estimation}. For continuously-defined treatment levels, \citet{hirano2004} proposed a direct translation of the propensity score by replacing the conditional probability mass function with the conditional density function of treatment assignment given covariates, known as a generalized propensity score (GPS); while this approach leads to as many propensity scores as there are levels of the treatment it uses only one single score at a time. Although \citet{imaiandvandyk2004} similarly found that the conditional density function of treatment assignment given covariates could serve as a propensity score, they noted potential limitations of this approach and suggested instead using the linear predictor in regression models or other summary statistic that are of finite dimension. When treatment assignment occurs over time as in the case where an individual chooses to receive PEG insertion or not, we must allow for the possibility of time-dependent confounding.  To this end,
let $X_t$ denote a set of $p$-dimensional time-dependent covariates at time $t$ and $\calX_t=\left\{X_s,~0\le s\le t\right\}$ denote the history of covariates up to time $t$. Then, the probability of treatment assignment at time $t$ given the covariate history up to time $t$ is
\begin{equation}\label{eq:density.tz}
f(t\mid\calX_t) = \lim_{\epsilon\rightarrow 0} \epsilon^{-1} P\left(t\le T < t+\epsilon \mid \calX_t\right),
\end{equation}
where $f(t\mid \calX_t) = h(t\mid X_t)\exp\left\{-\int_0^t h(s\mid X_s)\, ds\right\}$
and the hazard function is
\begin{equation}\label{eq:haz.tz}
h(t\mid X_t) = \lim_{\epsilon\rightarrow 0} \epsilon^{-1} P\left(t\le T < t+\epsilon \mid T\ge t, X_t\right).
\end{equation}
Because $h(t\mid X_t)$ uniquely parameterises $f(t\mid \cal{X}_t)$, either model~\eqref{eq:density.tz} or model~\eqref{eq:haz.tz} may be regarded as a legitimate treatment assignment model for continuous treatment with time-independent or time-dependent confounding \citep{li2001,lu2005}.

Of note, $f(t\mid \cal{X}_t)$ is a function of the entire covariate history $\calX_t$, whereas the hazard function $h(t\mid X_t)$ is a function of $X_t$ only.  This subtle, yet important difference can lead to difficulties when extending methods proposed by \cite{imaiandvandyk2004} and \citet{hirano2004} to time-dependent confounding via standard hazard modeling. In addition, both \citet{li2001} and \citet{lu2005} used the hazard function $h(t\mid X_t)$ as a GPS for matching which allows for balancing $X_t$ at the time of treatment in a matched set. However, they did not establish the strong ignorability of treatment assignment given their time-dependent GPS; this property does not hold if $Y$ is associated with $\calX_t$ rather than just $X_t$, in which case their proposed procedures may not lead to valid causal inference. Additionally, their proposed methods are only applicable to studies with data routinely collected at regular intervals, which is often not true in clinical registries.



We propose the propensity process to correct for confounding in observational studies by balancing the covariate history $\calX_t$. After the propensity process is estimated,
bias-corrected data analyses can be achieved through matching or stratification \citep{rosenbaumandrubin1983}. Establishing formally the theoretical properties of the propensity process for time-independent confounding requires different arguments than those presented in \citet{imaiandvandyk2004}.


\section{Methods}\label{sec:PS}

\subsection{Notation and Assumptions}

Our framework is constructed through potential outcomes \citep{rubin2005causal}.  For $t\in [0,L)$, we define $U_t=T\wedge t$ as the treatment time restricted to time $t$ and $U=T\wedge L$ as the treatment time restricted to time $L$, where $a \wedge b$ denotes the minimum of $a$ and $b$. Let $\calT_t=\left\{[0,t),t+\right\}$ define the set of potential treatment times restricted to $t$, $t\in[0,L)$, where $t+$ means that a patient did not receive PEG treatment before $t$. Let $Y^*_t$ be the potential outcome if a subject received PEG treatment at time $t$, $t\in[0,L)$, and $Y^*_{t+}$ the potential outcome if a subject did not receive PEG treatment in the interval $[0,t)$. It follows that $Y^*_{L+}$ denotes the potential outcome if a subject did not receive PEG treatment in the interval $[0,L)$.  We also define the treatment-free potential covariate process $\calX^*_t,~t\le L$. Then, the set of potential outcomes and treatment-free potential covariate process for a randomly selected subject from the population is $\{Y^*_s, \calX^*_s,~s\in \calT_t\}$ when treatment time is restricted at $t$, $t\in[0,L)$. In contrast, the observed data are $(Y, U, \calX_{U})$, where the observed outcome $Y=Y^{*}_{U}$, and the observed covariate history $\calX_{U}=\calX^*_{U}$.

Given $\theta_t=h(t\mid X^*_t)$, we define the propensity process as the sample path of the hazard function from baseline to time $t$, i.e.,
\begin{equation}\label{eq:pp}
\Theta_t = \left\{ \theta_s=h(s\mid X^*_s),~0\le s\le t \right\},
\end{equation}
noting that $\Theta_t$ is dependent on $\calX^*_t$. As  $\calX^*_t$ is observable only up to $U$,  $\Theta_t$ is estimable only up to $U$. While this concept seems similar to the propensity function \citep{imaiandvandyk2004},  the distinguishing factor of the propensity process is that $\Theta_t$ depends on $t$ and is of infinite dimension and $\Theta_L$ cannot be fully estimated for subjects receiving PEG before $L$, whereas the propensity function in \citet{imaiandvandyk2004} only allows for incorporation of time-independent covariates and can be estimated for all subjects.

In our framework, we make two assumptions.
\begin{assump} [Stable unit treatment value assumption] The distributions of potential outcomes for different subjects are independent of one another.
\end{assump}
\begin{assump}[Strong Ignorability]  For every $t\in [0, L)$, $\mbox{pr}(U_t\in\calA\mid Y^*_s,\calX^*_t)=\mbox{pr}(U_t\in\calA\mid \calX^*_t)$ and $\mbox{pr}\left(U_t\in\calA\mid \calX^*_t\right)>0$ for all $s\in \calT_t$, $\calX^*_t$, and  $\calA\subseteq\calT_t$.
\end{assump}
Assumption~1 is a common assumption in causal inference. However, our Assumption~2 is defined for each time point $t$ and differs from the standard strong ignorability of treatment assignment assumption used in earlier work for balancing scores. One implication of Assumption~2 is that, conditional on the treatment-free history $\calX^*_t$, receiving treatment at $t$ or not is independent of the set of potential outcomes, allowing us to model treatment assignment without conditioning on potential outcomes. 

\subsection{Main results}\label{theory}

We establish the large-sample results of the propensity process assuming that the true propensity process is known along the lines of \citet{rosenbaumandrubin1983} and \citet{imaiandvandyk2004}.
\begin{prop}
\label{prop1}
$U$ is conditionally independent of treatment-free covariate history $\calX^*_L$ given $\Theta_L$, where $\calX^*_L$ and $\Theta_L$ are the entire treatment-free covariate history and propensity process, respectively.
\end{prop}
Proposition 1 establishes $\Theta_L$ as a balancing functional that balances the entire covariate history. Proposition 1  requires that  $\Theta_L$ is known or can be estimated in the entire domain $[0,L)$.  In practice, however, we can only observe the covariate process $\calX^*_{U}$ and hence estimate $\Theta_{U}$. Proposition 2 establishes the balancing property for every given time point $t$ in $[0,L)$.

\begin{prop}\label{prop2}
For every $t\in[0,L)$, $U_t$ is conditionally independent of treatment-free covariate history $\calX^*_t$ given $\Theta_t$, where $\calX^*_t$ and $\Theta_t$ are the treatment-free covariate history and propensity process through time $t$, respectively.
\end{prop}
When $t=U$ in Proposition~\ref{prop2}, we have that  $U$ is independent of treatment-free covariate history $\calX^*_U$ given $\Theta_U$, where $\calX^*_U=\calX_U$ is observable and hence $\Theta_U$ is estimable. 
\begin{thm}\label{Thm1} For every $t\in [0, L)$, $\mbox{pr} \left(U_t\in\calA\mid Y^*_s,\Theta_t\right)=\mbox{pr}(U_t\in\calA\mid \Theta_t)$ for all $s\in \calT_t$, $\Theta_t$, and  $\calA\subseteq\calT_t$.
\end{thm}
When $t=U$ in Theorem~\ref{Thm1}, we have that  $U$ is independent of potential outcomes given $\Theta_U$, where $\Theta_U$ is estimable. Several remarks are in order. First, in \S~\ref{ssec:pp} we suggest modeling the hazard function in~\eqref{eq:haz.tz} through the proportional hazards model ~\eqref{eq:PH}; one could also use other model formulations for~\eqref{eq:haz.tz} and the results in Propositions~1--2 and Theorem~1 would still apply. Second, Proposition~\ref{prop2} and Theorem~\ref{Thm1} provide justifications for matching a subject treated at $t$ with an eligible control subject untreated at $t$ based on the propensity process up to $t$. It follows that each matched pair would have the same distribution for the covariate process up to $t$ and their potential outcomes are independent of their treatment assignments, allowing for valid causal inference. Third, our Proposition 2 is similar in spirit to Proposition 1 in \citet{lu2005} but is more general in the sense that the propensity process balances the entire covariate history up to $t$ not just the covariates measured at $t$. In addition, \citet{lu2005} did not establish the strong ignorability of treatment assignment given propensity scores similar to our Theorem 1. Proofs for Propositions~1--2 and Theorem~1 are given in the Appendix.

\section{Implementation and Practical Considerations} \label{methods}

\subsection{Interpolated Propensity Processes}\label{ssec:pp}

In practice, the propensity process $\Theta_{U}$ must be estimated from the observed data. The challenge for estimating the propensity process is that we may not observe the complete treatment-free covariate process $\calX^*_U$ on $[0,U]$; rather, we only get to observe the covariate process at a coarse set of discrete time points as is the case in the motivating ALS study.  Here, we propose to borrow strength across subjects in the study sample by modeling each time-dependent covariate as a random curve over time via nonlinear mixed effects models. This allows a predictive curve to be estimated for the entire treatment-free covariate process for each subject.

First, suppose we parameterize the hazard function in \eqref{eq:haz.tz} through Cox's proportional hazards model and define the propensity process
through the linear predictor,
\begin{align}\label{eq:PH}
&h(t\mid X_t;\beta)=h_{0}(t)\exp(\beta^{\rm T}X_t),
&&\Theta_t = \left\{ \theta_s=\beta^{\rm T}X^*_{s},~0\le s\le t \right\},
\end{align}
where $h_{0}(t)$ is the unspecified baseline hazard function.  Next, write the observed treatment-free covariate history for the $i$-th subject and $k$-th covariate as $\calX_{ik}=\left(X_{i1k},\ldots,X_{im_ik}\right)$, with time-dependent covariate $X_{ijk}$ measured at time $t_{ij}$. We note that the observation times $(t_{ij},~j=1,\ldots,m_i)$ may be different for each subject but are assumed to be the same for all covariates within a subject. Then, for each time-dependent covariate, we fit the model,
\begin{eqnarray}
X_{ijk} &=& b_k^{\rm T}(t_{ij})\gamma_k + b_k^{\rm T}(t_{ij})\alpha_{ik} + \epsilon_{ijk},~(i=1,\ldots,n;~j=1,\ldots,m_i;k=1,\ldots,p),\label{eq:mm}
\end{eqnarray}
where $\epsilon_{ijk}$ are independent, mean-zero random errors. To provide greater flexibility in modeling the covariate process over time, we use spline-type models \citep{ruppert2003semiparametric} in \eqref{eq:mm} where $b(\cdot)$ denotes a set of basis functions and $\gamma_k$ and $\alpha_{ik}$ are regression coefficients corresponding to the basis functions for the fixed and random effects, respectively. The interpolated treatment-free $\widehat{\calX}_t$ can be obtained from model~\eqref{eq:mm} by replacing regression coefficients $\gamma_k$ and $\alpha_{ik}$ with their estimates $\widehat\gamma_k$ and $\widehat\alpha_{ik}$, respectively. Then the estimated propensity process $\widehat\Theta_{U}$ can be obtained from~\eqref{eq:PH} by plugging in the interpolated $\widehat{\calX}_U$ and $\widehat\beta$, where $\widehat\beta$ is the estimated regression coefficient vector in the Cox proportional hazards model.


\subsection{Matching}\label{subsec:matching}


The use of matched analyses based on propensity scores for testing causal null hypotheses has been advocated by several other authors; for example, see  \citet{rosenbaumandrubin1983}, \citet{li2001} and \citet{lu2005} and references therein. Matching can be performed by minimizing the integrated squared error between the estimated propensity process $\widehat\Theta_{t}$ of a subject who received PEG treatment at time $t$ and that of each eligible control with $U>t$.  To accomplish this task, we implement a sequential matching algorithm. We start by ordering chronologically subjects according to their time of PEG treatment or censoring, namely $U$.  Set the matched pair counter to $m=1$ and select the subject with the smallest time to PEG treatment, say subject $i_1$.  Define the integrated squared difference in interpolated propensity processes between $i_1$ and $l$ as $Q(i_1,l) = I(T_{i_1}\le L) \int_0^{T_{i_{1}}} (\widehat\theta_{i_{1},t}-\widehat\theta_{l,t})^2\, dt,$ for all subjects $l$ in the set of $n-1$ eligible controls $\mathcal{C}_1=\{l\mid l=1,\ldots,n,~l\ne i\}$.  The matched control for $i_1$ is the nearest neighbor in interpolated propensity processes among eligible controls, i.e., $\mbox{argmin}_{l\in\mathcal{C}_1} Q(i_1,l)$. Increment the matched pair counter by one to $m=2$ and select the subject with the smallest time to PEG treatment, say $i_2$, excluding the two subjects in the first matched pair.  Therefore, the set of eligible controls, say $\mathcal{C}_2$, contains $n-3$ subjects: all $n$ subjects less the two subjects in the first matched pair and $i_2$. The matched control for $i_2$ is the nearest neighbor in interpolated propensity processes among the set of eligible controls, $\mbox{argmin}_{l\in\mathcal{C}_2} Q(i_2,l)$.  Increment the matched pair counter by one and continue until all treated individuals are matched or until there are no suitable controls available for matching.

\section{Analysis of the ALS Registry Data} \label{results}

Using a data set from the Emory ALS Registry, we assess the association of PEG treatment with the change in body mass index (BMI) from baseline to 18 months, i.e., $L$ = 18 months. The data set includes 240 patients who survived past $L$ and had at least one clinic visit between baseline and $L$. The patients who received PEG did so after their first clinic visit. The timing of recommending PEG by the physician involved many factors and the final decision to have PEG was made by each patient. We model treatment
assignment through the proportional hazards model \eqref{eq:PH} including the following covariates. The baseline risk factors are age at diagnosis, sex, site of onset of disease, negative inspiratory force, and time from diagnosis to the first clinic visit.  Two time-varying covariates are forced vital capacity and body mass index, which may not be measured at every clinic visit for every patient. Each time-varying covariate is modeled over time using the mixed model~\eqref{eq:mm}, where polynomial spline basis functions are used. The estimated curves are used to interpolate the covariate values needed for estimating the propensity process based on \eqref{eq:PH}.

We compare three alternative approaches to the proposed propensity process. First, a na\"{i}ve analysis compares all treated individuals to those who are untreated prior to $L$. The second approach is the propensity function \citep{imaiandvandyk2004} that uses baseline risk factors $X_0$ only in the treatment assignment model~\eqref{eq:PH}, where $\theta_0=\beta^{\rm T}X_0$ defines the propensity function. The third approach is the interpolated generalized propensity score, which uses the interpolated treatment-free $\widehat{X}_t$ defined in \S~\ref{ssec:pp} to obtain the GPS for each subject in the spirit of \citet{lu2005}, noting that $X_t$ may not be observed at time $U$ for a subject and its eligible controls as defined in \S~\ref{subsec:matching}. The same sequential matching algorithm in \S~\ref{subsec:matching} is used for all propensity score methods.  Our matching algorithm resulted in $M=74$ pairs for the analysis using the propensity function and $M=76$ pairs for both analyses using the generalized propensity score and propensity process.

Following \citet{li2001} and \citet{lu2005}, we assess balance of covariates by examining Type I errors from a log-rank test of the effect of the covariate on time to treatment, one covariate at a time. In the matched analyses, this model is stratified by the $M$ matched pairs. As shown in Table 1, prior to matching, balance is not achieved. While other methods improve covariate balance,  they do not balance all covariates. However, matching using the propensity process results in balance across all covariates. This indicates that the propensity process outperforms the baseline propensity function or interpolated GPS in terms of balancing covariates and there may be residual confounding after matching by the other propensity score methods.

\begin{table}[h]\label{tab:balance}
  \centering
  \caption{Covariate balance before and after matching}
    \begin{tabular}{lcccc}
 \hline
             & Prior to  & Propensity   & Generalized  & Propensity      \\
    Covariate &         Matching        & Function & Propensity Score & Process \\\hline
    Body mass index & 0.277 & 0.245 & 0.986 & 0.991 \\
    Forced vital capacity & 0.764 & 0.539 & 0.201 & 0.317 \\
    Negative inspiratory force & 0.151 & 0.022 & 0.016 & 0.704 \\
    Age & 0.162 & 0.718 & 0.378 & 0.195 \\
    Sex & 0.577 & 0.695 & 0.002 & 0.706 \\
    Site & 0.001 & 0.003 & 1.000 & 0.341 \\
    Time from diagnosis & 0.676 & 0.633 & 0.033 & 0.854\\
\hline
    \end{tabular}
\end{table}

After matching, we test the causal null hypothesis that the mean potential outcome is the same whether a patient received PEG treatment at time $t$ versus PEG treatment at some time after $t$ or untreated by $L$, which can be written as $H_0: E(Y^*_t)=E(Y^*_s)$ for all $t<s\leq L$. We test this hypothesis by a Wilcoxon signed rank test on matched pairs for all the matched analyses.  The Wilcoxon rank sum test is used for hypothesis testing in the na\"ive analysis.  Table 2 presents the median difference in BMI change at 18 months and p-value of the Wilcoxon test for each approach. The propensity process matched analysis suggests a protective effect of PEG on BMI, whereas the other three methods all show effects that are attenuated towards 0 and are not statistically significant.

\begin{table}[h]\label{tab:outcome}
  \centering
  \caption{Results in the data analysis}
    \begin{tabular}{lcc}
    \hline
       & Median Difference&  P-value\\
    \hline
    Na\"ive & 0.035 & 0.673 \\
    Propensity Function& 0.030 & 0.466 \\
    Generalized Propensity Score& 0.360 & 0.453 \\
    Propensity Process& 0.830 & 0.022 \\
    \hline
    \end{tabular}
\end{table}

\section{Discussion} \label{discussion}
Compared to the existing propensity score methods, the propensity process offers the advantage of balancing time-varying covariate history from baseline to time of treatment. A key  component of this approach is the interpolation of covariate curves. We propose to use nonlinear mixed models to provide flexibility for modeling covariate history, though there must be enough individual longitudinal data collected to estimate these curves, which is a potential limitation in settings with sparsely collected longitudinal data. However, data interpolation may not be needed in settings such as critical care in intensive care units where time series data including heart rate and blood pressure are continuously recorded \citep{lehman2013tracking}.


In our data analysis, we use a straightforward approach for hypothesis testing after matching. Future extensions may include conditional likelihood methods for estimating treatment effects based on matched pairs/sets and methods for stratification and covariate adjustment using the propensity process. Additionally, our analysis excludes individuals who died prior to $L$ in order to avoid complications due to censoring by death \citep{rubin2006,zhang2003estimation}, which could be addressed in future extensions such that no such exclusion is necessary.


\section*{Acknowledgements}
The authors thank Dr. Jonathan Glass and Ms. Meraida Polak at the Emory ALS Center for providing the ALS data and Dr. Xin Qi at the Georgia State University for helpful comments.

\appendix
\section*{Appendix: Proofs of Propositions 1, 2, and Theorem 1}

\noindent\textit{Proof of Propositions 1 and 2:}  We prove Propositions 1 and 2 based on the treatment assignment model defined in \eqref{eq:density.tz} and \eqref{eq:haz.tz}. Given $\theta_t=h(t\mid X^*_t)$,
\begin{eqnarray}
f(t\mid \calX^*_t,\Theta_t)&=&f(t\mid \calX^*_t)\nonumber\\
                           &=& h(t\mid X^*_t)\exp\left\{-\int_0^t h(s\mid X^*_s)\, ds\right\}\nonumber\\
                           &=& \theta_t\exp\left\{-\int_0^t \theta_s\, ds\right\}\label{eq:f}\\
                           &=& f(t\mid \Theta_t), \mbox{ for all $t \in [0,L)$},\nonumber
\end{eqnarray}
where the first equality is due to the fact that $\Theta_t$ is redundant given $\calX^*_t$. It follows from integrating
both sides in $[0,L]$ that $\mbox{pr}(T\geq L\mid \calX^*_L,\Theta_L)=\mbox{pr}(T\geq L\mid \Theta_L)$. The result in Proposition~1 follows immediately, i.e., $U$ is conditionally independent of $\calX^*_L$ given $\Theta_L$. Along similar lines, we can prove the result in Proposition~2, i.e., $U_t$ is conditionally independent of $\calX^*_t$ given $\Theta_t$ for all $t \in [0,L)$.

\medskip
\noindent\textit{Proof of  Theorem 1:} For every $t \in [0,L)$, all $s\in \calT_t$, $\Theta_t$, and $\calA\subseteq\calT_t$,
\begin{eqnarray}
\mbox{pr}\left(U_t\in\calA\mid Y^*_s,\Theta_t\right)&=&E\left\{\mbox{pr}\left(U_t\in\calA\mid Y^*_s,\calX^*_t\right)\mid Y^*_s,\Theta_t\right\}\label{eq:prop3.1} \\
                                        &=&E\left\{\mbox{pr}\left(U_t\in\calA\mid \calX^*_t\right)\mid Y^*_s,\Theta_t\right\}\label{eq:prop3.2}\\
                                        &=&E\left\{\mbox{pr}\left(U_t\in\calA\mid \calX^*_t,\Theta_t\right)\mid Y^*_s,\Theta_t\right\}\label{eq:prop3.3}\\
                                        &=&E\left\{\mbox{pr}\left(U_t\in\calA\mid \Theta_t\right)\mid Y^*_s,\Theta_t\right\}\label{eq:prop3.4}\\
                                        &=&\mbox{pr}\left(U_t\in\calA\mid \Theta_t\right). \label{eq:prop3.5}
\end{eqnarray}
Let $\sigma(Y^*_s,\Theta_t)$ and $\sigma(Y^*_s,\calX_t)$ denote the $\sigma$-field generated by $(Y^*_s,\Theta_t)$ and $(Y^*_s,\calX_t)$, respectively. By the definition of $\Theta_t$ in~\eqref{eq:pp}, we have $\sigma(Y^*_s,\Theta_t)\subseteq\sigma(Y^*_s,\calX_t)$ and then \eqref{eq:prop3.1} follows immediately \citep[cf.][Theorem 34.4]{billingsley2008probability}. \eqref{eq:prop3.2} is due to Assumption~2 while \eqref{eq:prop3.3} follows from the fact that $\Theta_t$ is redundant given $\calX^*_t$. The fourth expression \eqref{eq:prop3.4} is due to Proposition~2 while \eqref{eq:prop3.5} follows from~\eqref{eq:f}, i.e., $\mbox{pr}\left(U_t\in\calA\mid \Theta_t\right)$ is independent of $Y^*_s$.

\bibliographystyle{apa}
\bibliography{PP}

\end{document}